\documentstyle[12pt,epsf]{article} 

\begin{document}

\begin{flushright}
{\normalsize SLAC--PUB--7551}\\
{\normalsize DOE/ER/40561--329--INT97--21--05}\\
{\normalsize DOE/ER/41014--15--N97 }\\
\hspace {3.7in} August 1997\\
\end{flushright}
\centerline{\Large\bf Chiral Limit of Nuclear Physics}
\bigskip

\centerline{Aurel BULGAC$^a$ \cite{AB},
Gerald A. MILLER$^{a,b,c}$ \cite{GAM} and Mark STRIKMAN $^{b,d,e}$ \cite{MS}
\vspace{15pt} }

\centerline{$^a$ Department of Physics, P.O. Box 351560}\vspace{5pt}
\centerline{University of 
Washington, Seattle, WA 98195--1560 \vspace{15pt}}

\centerline{$^b$ Institute for Nuclear Theory, P.O. Box 351550}\vspace{5pt}
\centerline{University of 
Washington, Seattle, WA 98195--1550 \vspace{15pt}}

\centerline{$^c$ Stanford Linear Accelerator Center, Stanford University}
\vspace{5pt}
\centerline{
Stanford, CA 94309\vspace{15pt}}

\centerline{$^d$
Pennsylvania State University, University Park, PA 16802 \vspace{15pt}}
\centerline{$^e$ 
Institute for Nuclear Physics, St. Petersburg, Russia} 

\begin{abstract}
We study nuclear physics in the chiral limit ($m_u$, $m_d=0$) in which 
the pion mass vanishes. We find that the deuteron mass is 
changed little, but that P--wave 
nucleon--nucleon scattering volumes are
infinite. This motivates an investigation of the  possibilities that there 
could be a  
two--nucleon $^3P_0$ bound state,  and that 
the nuclear matter ground state is likely to be a condensed state 
of nucleons paired to those  quantum numbers. However, 
the short distance repulsion in the nucleon--nucleon 
potential is not affected by the chiral limit and prevents such 
new chiral possibilities. 
Thus the  chiral limit physics of nuclei is very similar to that
of nature. Using the  chiral limit to simplify  QCD sum rule calculations
of  nuclear matter properties seems to be a reasonable approximation.
\end{abstract}

\newpage

The derivative coupling of pions to nucleons, which results from
spontaneously broken chiral symmetry, plays a major role in the
structure of nuclei. This coupling, suppressed by the small momenta
typical of nuclear physics, prevents the nuclear ground state, a system
of strongly interacting hadrons, from being a pion soup. The net result
is that the pion dominated internucleon interactions play a significant,
but not dominant, role in the structure of the nuclei. The effects of
the two pion exchange potential are believed to account for some of 
the needed
mid--range attraction, but are not strong enough to induce a phase
transition from the normal Fermi liquid that is believed to describe the
ground state of heavy nuclei.

During recent years there has been a renewed interest in applying 
chiral Lagrangian models and QCD sum rules towards the description of 
microscopic nuclear structure\cite{lr,cohenrev}. An important element of
these analyses is the implicit assumption that the chiral limit $m_q
\rightarrow 0$, in which $m_{\pi} \rightarrow 0$ as well, is smooth
enough and does not change qualitatively properties of nuclei. This is
reflected in an assertion that the major nuclear characteristics can be
expressed through various vacuum condensates which can be determined in
the $m_q \rightarrow 0$ limit. In particular, it was suggested that in
this limit one can use the $\omega, \sigma$--model where pion degrees of
freedom are essentially irrelevant\cite{revs}. 

This assumption raises a question: 
What is the nature of nuclei in the $m_q
\rightarrow 0$ limit? One can see immediately that the one pion exchange
potential acquires a long range tensor interaction between nucleons $
\propto r^{-3}$. This new interaction raises the possibility that in
the chiral limit the structure of the nuclear ground state may be very
different from that of actual nuclei. In the absence of quark masses
 the only scale in
QCD is $\Lambda_{QCD}\approx 200$  MeV, so that one might naively
expect that the binding energy of the deuteron and the binding energy
per nucleon in nuclear matter might be of that order of magnitude.

The purpose of the present note is to analyze 
nuclear properties in the chiral limit. This is based on  the
chiral limit of the nucleon--nucleon $NN$ interaction. To take this limit
we need to analyze the different scales. Our view is that
the $NN$ interaction can be
understood in terms of two mass scales: 
the $m_\pi$ scale which governs the long range physics and 
the $m_{\rho}$ scale which governs short range physics.
 We explicitly set $m_\pi$ to zero in the long range
part of the nucleon--nucleon potential, but assume that the
short distance physics is
unaffected by taking the chiral limit.
Thus the short range $NN$
interaction is taken to be  the
same as for realistic $NN$ interactions.
Consider, for example, the Bonn
one boson exchange model\cite{bonn}, in which the short 
range part of the potential is essentially due to the
exchange of $\omega$, and $\rho$ mesons. Our assumption
is that the masses and the
interaction strengths of the vector mesons are not changed in the chiral
limit. 

The effects of the chiral limit on the medium range attraction 
are more subtle. It is typical (as is the case in Ref.\cite{bonn})
to account for this physics by using
the exchange of a $\sigma$ meson of mass 550 MeV. 
Here we shall assume that the exchange of a $\sigma$ meson is not influenced
by taking the chiral limit, even though this 
could also be a phenomenological method of 
including the contribution of the exchange of uncorrelated 
pions to the two pion exchange potential. In this case, one would expect
that taking the  chiral  limit would influence the medium range 
attraction. We do not investigate 
such effects here and our conclusions are based  on 
setting $m_\pi$ to zero in the one
pion exchange potential (OPEP).
Note also that contribution of correlated two pion exchange due to 
excitation of a $\Delta$ isobar in the intermediate state is also not
affected significantly by the change of the pion mass since the range
of this interaction is mainly determined by the mass difference of the
nucleon and $\Delta$ which practically does not depend on $m_q$.

We proceed by arguing that the 
the value of the pion--nucleon coupling constant $g$  should be little
affected by taking in the chiral limit. Indeed, in this limit the
Goldberger--Trieman relation ($g/2m=g_A/2f_\pi$) would hold exactly and
the coupling constant would change only if the nucleon mass $m$ and the
pion decay constant $f_\pi$ were to depend on the pion mass. The
quantity $f_\pi$ depends on the wave function of the $q\bar q$ component
at the origin, which is expected to be essentially independent of the value
of $m_\pi$. Similarly, the nucleon mass $m$ is not expected
to be significantly different in the chiral limit, see Ref.
\cite{cohen96}. The value of $m_\pi^2$ enters, via chiral loop graphs, as
a term of the form $m_q \ln (m_q$), which vanishes in the chiral limit.
We thus conclude that the nucleon mass $m$
and $f_\pi$ are not significantly changed by considering the chiral
limit, and that $g$ should remain close to its measured value. 

Thus in the chiral limit $m_\pi \rightarrow 0$ the OPEP becomes:
\begin{eqnarray}
 V(\vec{r})&=&-{g^2\over 4\pi} {1\over 12m^2} 
\vec{\tau_1}\cdot\vec{\tau}_2
\vec{\sigma}_1\cdot
 \vec{\sigma}_2 {1\over 12\;\pi} 
{\Lambda^3\over 2}
 \exp(-\Lambda r)\nonumber\\
 &+& {g^2\over 4m^2} \vec{\tau_1}\cdot\vec{\tau}_2
 S_{12}{1\over 4\pi}
 \left [ \frac{1}{r^3}-\left (
\frac{1}{r^3}+\frac{\Lambda}{r^2}+\frac{\Lambda^2}{2 r} +
\frac{\Lambda^3}{6}\right ) \exp (-\Lambda
 r) \right ] \label{full}
\end{eqnarray}
where the parameter $\Lambda$\cite{bonn}
governs the falloff of the monopole 
$\pi$--nucleon form
factor. For values of $r$ such that $r\gg 1/\Lambda$ one is left with a
tensor term that  falls off rather slowly as $1/r^3$.

Does the chiral limit affect the deuteron?
{}From Eq.~(\ref{full}) we see that the long range attraction present in the 
central potential is absent in the chiral limit, and that the attractive
tensor force is enhanced. We compute the change in the deuteron binding energy 
by using first order perturbation theory and find  a reduction of 
about 1 MeV. Specifically, if the deuteron  wave function of the Paris
NN  potential\cite{paris} is
used, 
about 2 MeV of attraction (from the central part of the OPEP)
is lost, and 0.9 MeV  is gained (from the tensor part of the OPEP), 
which is a loss
of 1.1 MeV of binding.
For the Bonn potential,  the numbers are 2 MeV and $-1.3$ MeV, or a loss of
about 0.7 MeV. This small change seems inconsequential.

We next turn to the $P$--waves.
The scattering volume is a measure of the $P$--wave scattering at low
energies. As discussed by Ericson and Weise \cite{ew} this volume is
qualitatively reproduced by computing the scattering amplitude from the
OPEP in first order Born approximation.

In the chiral limit (and ignoring the effects of the form factor)
the spin--triplet $P$--wave Born amplitudes are given
by 
\begin {equation}
 \left[{ e^{i\delta}sin\delta\over p}\right]_{^3P_J} =-{g^2\over 16\pi M}
 <L=1,S=1,J|S_{12}|L=1,S=1,J>\int r^2dr {j_1^2(pr)\over r^3},
\end {equation}
in which we keep only the long ranged term of Eq. (\ref{full}) and where
$p$ is the relative momentum, $<L,S,J|S_{12}|L,S,J>=\{-(2L+2)/(2L-1),2,
-2L/(2L+3)\},$ for $J=\{L-1,L, L+1\}$, or $\{-4,2,-4/5\}$ for $L=1$.
The radial integral is dimensionless and it equals 1/4.
Thus for
$p\rightarrow 0$, the phase shift behaves as $\delta(^3P_J) \propto p$
and the scattering volume $a(^3P_J)$, defined by the relation 
$a(^3P_J)=\lim_{p\to 0} \delta(^3P_J)/ p^3$ is infinite in the
chiral limit. 

This behaviour of the phase shift and the infinity of $a(^3P_J)$ is not
an artifact of either retaining only the longest ranged term of Eq.
(\ref{full}) or using the first order Born approximation. Any other
shorter range component of the $NN$ interaction potential leads to a
contribution to the phase shift proportional to $p^3$. One can also
consider very small momenta $p$, for which the phase shift is very small
and thus the first order Born approximation holds.

Even though we have discussed only $P$--waves, one can show that for all
non--zero angular momentum waves the phase shifts have a qualitatively
similar behaviour, i.e. $\delta(^3L_J) \propto p$. This means that 
there is significant scattering at long range for all partial waves.
Another way of showing this is by considering the first order Born
approximation to the OPEP scattering amplitude $f_B$
\begin{equation}
f_B\propto 
{3\vec{\sigma_1}\cdot \vec{q} \vec{\sigma_2}\cdot \vec{q}
\over \vec{q}^2+m_\pi^2},
\end{equation}
where $\vec{q}$ is the momentum transfer. The quantity $f_B$ vanishes
in the forward direction $\vec{q}=0$. Taking the chiral limit of a 
$m_\pi \to 0$, causes a non--vanishing amplitude which is independent
of the magnitude of $\vec{q}$, and therefore contains significant
scattering at all partial waves.

In the $^3P_0$
channel the expectation value of $S_{12}$ 
has the relatively large magnitude of $\;4$, and the OPEP is attractive.
This causes us to examine the possibility that  the OPEP interaction
in the chiral limit could  lead to a bound state. 
In this channel the tensor potential
is given by 
\begin{eqnarray}
 V(r,^3P_0)=-{g^2\over 4\pi\;m^2} 
 \left[ \frac{1}{r^3}-\left ( 
\frac{1}{r^3}+\frac{\Lambda}{r^2}
+\frac{\Lambda^2}{2 r} +\frac{\Lambda^3}{6}\right )
\exp (-\Lambda
 r)\right]. \label{3p0}
\end{eqnarray}
The effective potential, $V_{eff}$,
    defined by adding the centrifugal barrier
of $2/(m\;r^2)$ to the 
potential of Eq.~(\ref{3p0}) is shown in Fig.~1 (for $\Lambda=1.8$ 
GeV).
 This quantity has a
broad minimum which is deep enough to support a bound state. 
However, one must also include the effects of the short range
potential. Fig.~1 shows the effects of using the 
Argonne V18 potential\cite{v18} to represent the short distance
interaction. This repulsive core causes the minimum to be eliminated
and there will be no  bound state. 

\begin{figure}[htbp]
\centerline{\epsffile{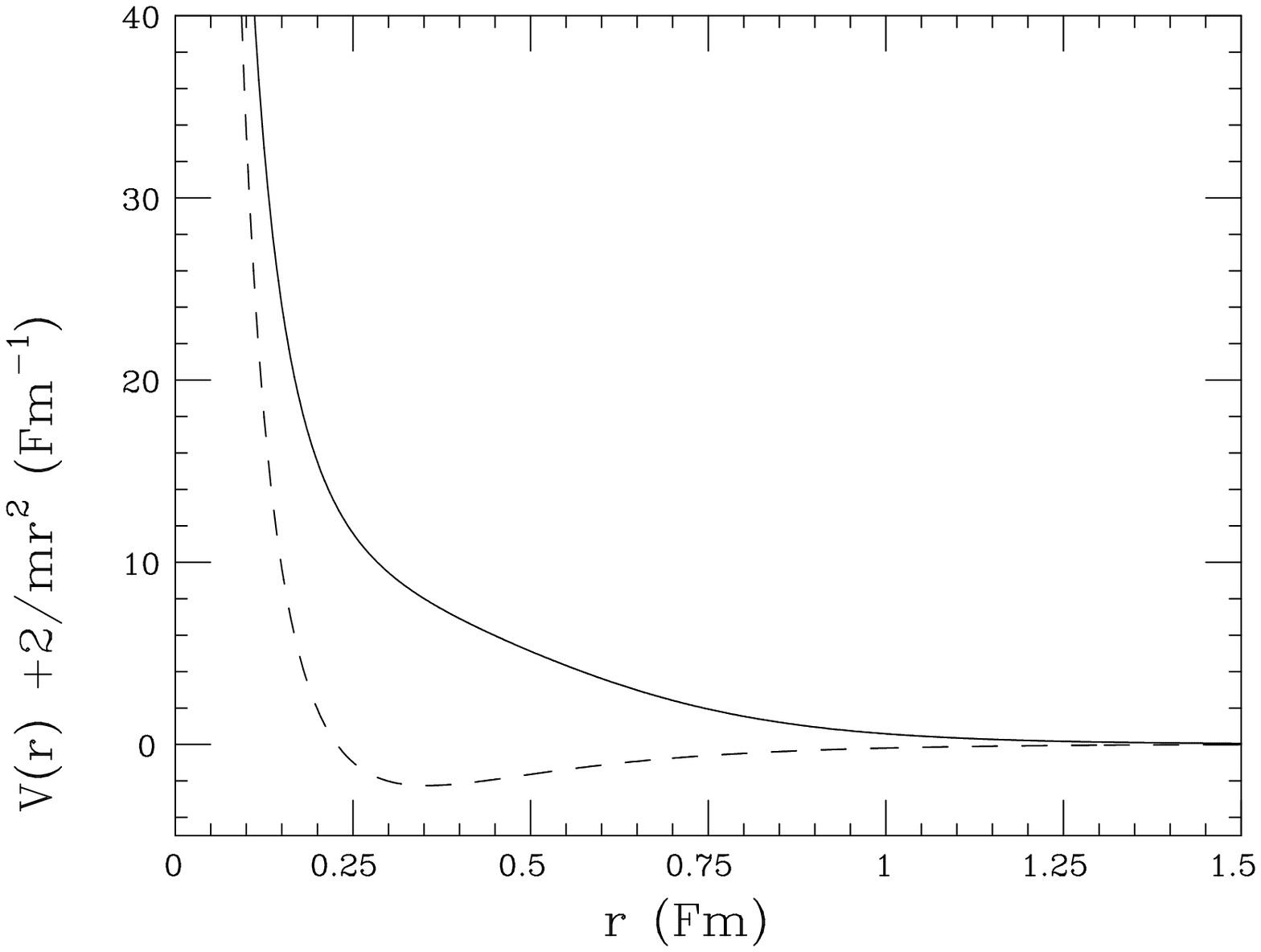}}
\caption{$V_{eff}=V(r)+2/mr^2$ for the $^3$P$_0$ channel. 
Dashed curve -- $V(r)$ from Eq.~(\ref{3p0}). Solid curve -- includes also the
short range part of the NN  potential.}
\end{figure}

Even if no bound $NN$ state with quantum numbers $T=1$ and $^3P_0$
exists, the chiral  limit 
interaction in this channel is significantly stronger
and has a significantly longer range that the usual
OPEP potential. Thus we consider consequences for nuclear physics.
 In nuclear matter only the short range part of the $NN$
interaction is strongly renormalized \cite{geb,bethe} (the healing
length is rather short), while the long range part survives. Therefore, 
the only qualitatively new element is the possibility to
form nuclear matter with pairing properties analogous to bulk $^3$He
\cite{he3}, namely $0^-$ nucleon pairs. $^3P_0$ pairing is in some
respects analogous to usual $^1S_0$ pairing \cite{he3}, as it leads to a
isotropic momentum space gap and the condition for its appearance is
formally similar to the usual BCS pairing gap equation. 

We take up the notion of this new kind of condensation 
for infinite nuclear matter. For that system the gap equation
is given by 
\begin{eqnarray}
\Delta(k)=-\int_0^\infty\;p^2dp 
{\langle k|V(^3P_0)|p\rangle\Delta(p)\over 
2\sqrt{({p^2\over 2M}-\varepsilon_F)^2+\Delta^2(p)}},
\label{gap}
\end{eqnarray}
where $\langle k|V(^3P_0)|p\rangle$
 is the momentum--space version of the potential
of Eq.~(\ref{3p0}). The use of the interaction of Eq.~(\ref{3p0}),
leads to 
nontrivial solutions for a range of
values of $\Lambda$.
The value $\Delta(k)$ is typically 
several MeV for  $k\approx k_F$ for $\Lambda\approx 1000$ MeV/c and
significantly
larger for larger values of $\Lambda$. 
It is amusing that in the chiral limit nuclei could
become unstable with respect to the formation of a condensate of
$(T,J^\pi)=(1,0^-)$ pairs, with the same quantum numbers as the real
pions, which lead to this instability. 

However, including the effects
 of the short ranged repulsive interaction causes the nontrivial 
solutions to
 be  eliminated, such that $\Delta(k)=0$.

We also examine some other situations for which the chiral limit has little
impact. 
First we take up 
question of whether the increased attraction is sufficient to cause
nuclei to condense into a crystalline solid\cite{did}.
This would occur for
potentials that are strong enough to overcome the zero--point
motion of nucleons about a possible lattice site. However, in the
chiral limit the OPEP potential is not strong enough to lead to
additional bound
states. 
If more bound states than the deuteron would have
appeared, and if the short range repulsion were absent,
one could have expected that nuclear matter and nuclei in
particular would have had a crystalline structure in their ground state.
For potentials of the strength that we find, the zero point
oscillations would have an amplitude of the same order of magnitude as
the size of the deuteron, so we don't expect crystallization
in the chiral limit.

Another possibility to consider is the formation of a condensate of real
pions. In the chiral limit the pion mass is zero, and apparently there
is no energy cost to generate real pions and thus create a ``real pion
condensate'', which differs from the condensate of virtual pions
suggested by Migdal \cite{mig}.
 The energy penalty to create a zero mass pion, localized inside
a large nucleus is of the order of $\pi/R_A\approx 100$ MeV and thus a
``real pion condensate'' is extremely unlikely. We also find, by
explicitly
using $m_\pi=0$ in the usual equations\cite{ew,mig},
that the possibility of a Migdal condensate is not  enhanced in the
chiral  limit.

Finally, we consider the effects of the increased range of the tensor
interaction for the binding energy of infinite nuclear matter. This
enters in second--order in a non--relativistic calculation\cite{geb}.
Explicit evaluate of this shows that the effects are negligible.

The net result is that the physics of nuclei in a universe where
$m_\pi=0$ would be similar to what is actually observed.
The QCD sum rule calculations of nuclear matter
properties\cite{lr,cohenrev}
are not rendered erroneous through their use of the chiral limit.

This is work is supported  by the US DOE.  We thank G.F. Bertsch for
making a very useful suggestion.
We thank D.S. Koltun for a useful discussion.


\end{document}